\documentclass[11pt]{article}

\usepackage{graphicx}
\usepackage{textcomp}
\usepackage{amsmath}
\usepackage{amsfonts}
\usepackage{epsfig}

\newcommand{\be}{\begin{equation}}
\newcommand{\ee}{\end{equation}}
\newcommand{\beqn}{\begin{eqnarray}}
\newcommand{\eeqn}{\end{eqnarray}}
\newcommand{\beqns}{\begin{eqnarray*}}
\newcommand{\eeqns}{\end{eqnarray*}}

\newcommand{\lkr}{\left(}
\newcommand{\lkv}{\left[}
\newcommand{\rkv}{\right]}
\newcommand{\rkr}{\right)}
\newcommand{\lfi}{\left\{}
\newcommand{\rfi}{\right\}}

\newcommand{\fr}[1]{(\ref{#1})}

\newcommand{\af}{\alpha}
\newcommand{\eps}{\varepsilon}
\newcommand{\Ga}{\Gamma}
\newcommand{\ga}{\gamma}

\newcommand{\lam}{\lambda}

\newcommand{\sig}{\sigma}
\newcommand{\Lam}{\Lambda}

\newtheorem{theorem}{Theorem}
\newtheorem{lemma}{Lemma}
\newtheorem{corollary}{Corollary}

\newcommand{\EE}{\ensuremath{{\mathbb E}}}
\newcommand{\II}{\ensuremath{{\mathbb I}}}

\newcommand{\Tr}{\operatorname{Tr}}
\newcommand{\sign}{\operatorname{sgn}}
\newcommand{\pen}{\operatorname{pen}}
\newcommand{\diag}{\operatorname{diag}}

\begin{document}

\title
{\bf  De-noising procedures for frame operators  }

\author{
         {\em Daniela De Canditiis},\\
              Istituto  per le Applicazioni
               del Calcolo ``M. Picone'',
              CNR Rome, Italy,\\
               \\
         {\em Marianna Pensky},\\
         Department of Mathematics,
         University of Central Florida,\\
                    \\
       {\em Patrick J. Wolfe},\\
        Department of Statistical Science,
 University College, London  }

\date{}

\bibliographystyle{plain}
\maketitle

\begin{abstract}
The present paper provides a  comprehensive  study of   de-noising
properties of frames and, in particular, tight frames, which constitute one of the
most popular tools in contemporary signal processing.  The objective of the  paper is to bridge the
existing gap between mathematical and statistical theories on one hand and engineering practice on the other
and explore how one can take advantage of a specific structure of a frame
in contrast to an arbitrary collection of vectors or an orthonormal basis.
For both the general and the tight frames, the paper presents
a set of practically implementable de-noising techniques which take frame induced correlation structures
into account.  These results are supplemented by an examination of the case when the frame is constructed as
 a collection of orthonormal bases. In particular, recommendations are given for aggregation of the
estimators at the stage of frame coefficients. The paper is concluded by a finite sample simulation study
which confirms that taking frame structure and frame induced correlations  into account indeed improves de-noising precision.

\vspace{3mm} {\bf  Keywords}: {frames, tight frames, shrinkage, thresholding,aggregation }

\vspace{3mm}{\bf AMS (2000) Subject Classification}: {Primary 62G08;
Secondary 42C15}

\end{abstract}

\section {Introduction}
\label{sec:intro}
\setcounter{equation}{0}

\medskip

In the recent years,  there has been resurgence of interest in   de-noising by  frames  spanning different communities.
 The effort was undertaken by mathematicians working in the area of approximation theory,
by  statistics and computer science communities (the ``large-$p$,
small-$n$'' problem and model  selection),
and   by  engineering community (regularization theory and sparse coding
of signals and images).

The need for overcomplete representations stems from the fact that, though a single orthogonal basis
allows very fast computations, it very often fails to efficiently represent a function of interest, $f$,
so that one needs a large number of coefficients to transmit or store.   In fact, if $f$ is
expanded over a much more exhaustive dictionary with $p$ elements, it very often can be represented
with a very few nonzero coefficients. Moreover, one can reduce the error of representation beyond what is possible
when one orthonormal basis is used.

Mathematicians and statisticians dealt with this problem for years.  However,  the methods
which they designed were intended for an arbitrary dictionary  and did not take advantage
of their particular structure. For this reason, those methods work very well
in a regression-type set up   when  one does not need to obtain results instantaneously.

One of the most popular groups of methods relies on minimizing the difference between
the function $f$ and its representation under some set of constraints.
From the point of view of optimization theory
this problem can be re-formulated as the problem of minimization
of penalized risk of the representation of $f$. Various choices of penalties and
risk functions were suggested leading to RIDGE regression (see, e.g., Brown  and Zidek (1980),
BRIDGE regression (Frank  and Friedman (1993)), LASSO (Tibshirani  (1996)), Dantzig selector
(Candes   and Tao  (2007)), the least angle regression
(Efron  {\it et.al.} (2004) and Support Vector regression (Smola  and Sch\"{o}lkopf  (2004)) among others.

Such methods neither assume nor exploit any specific structure of the dictionary
and, as a result, are very computationally expensive. For this reason, those methods
cannot be used for real-time problems and, as a result,  are not very popular
in practical engineering applications.

For many years, engineers have been using frames, especially,
tight frames due to their simple reconstruction properties.
However, when it comes to de-noising, engineers routinely treat frames,
especially tight frames, as if they were  orthonormal bases
completely ignoring correlations between frame vectors
and using thresholding methodologies developed for the case of orthonormal bases.
 This sentiment is well expressed
in a recent paper of Yu,  Mallat  and Bacry (2008)  which states
that ``a tight frame behaves like a union of
$\af$  orthogonal bases.'' For this reason, there is a multitude of engineering papers
where methodologies designed for orthonormal basis are applied to frames without
any consideration of frame structure.

There have been a growing sentiment in the statistics community that
discounting correlations associated with frames reduce de-noising precision.
Few authors focus their attention on universal threshold which is frequently used by engineers in the context of frames
as if all frame functions are independent which leads to the threshold which is too large.
Among them, Downie and Silverman (1998) considered   multiwavelets which constitute  a particular type of a frame.
Walker and Chen (2010) studied   universal thresholding in the case of  Gabor frames with a Blackman window.
Recently, Haltmeier and Munk (2012) derived the universal threshold for a general frame
satisfying rather stringent conditions which ensure that the threshold depends on the number of frame functions but not
on frame structure.

However,  to the best of our knowledge, there have never been a   comprehensive  study of a de-noising
properties of frames and, in particular, tight frames, which constitute one of the
most popular tools in contemporary signal processing.
Note that general statistical methods designed
for correlated  noise do not lead to fast computations and are impractical in this set up
since the covariance matrix is too big. The objective of the present paper is to bridge the
existing gap between mathematical and statistical theories on one hand and engineering practice on the other
and explore how one can take advantage of a specific structure of a frame
in contrast to an arbitrary collection of vectors or an orthonormal basis.
In particular, the purpose of this paper is to provide a set of practically implementable
de-noising techniques which take correlation structure of the frame coefficients into account.

In pursuing this goal, we start with derivation of  the   oracle (best in the mean square sense) linear
diagonal shrinkage estimator. It turns out that, by construction, the i-th element of the diagonal shrinkage
matrix depends not only on the  i-th frame coefficient but also on coefficients related to it (not necessary in its vicinity).
In this sense, the oracle can be regarded as a block shrinkage procedure, where the length and the constitution of the
block is automatically determined by the  correlation structure of frame coefficients induced by the frame transform.
The oracle is followed by derivation  of  the Stein Unbiased Risk Estimator
(SURE) in the case of a general frame and an arbitrary de-noising strategy.
The SURE formulation is very similar to the one obtained by Blu and Luisier (2007)
for the case of interscale image de-noising.
Next, we use this result for designing particular types of de-noising algorithms
(linear shrinkage, soft or hard thresholding, etc.)
The   SURE  provides a good assessment tool for de-noising with any kind of a frame and can, in fact, be
used for construction of frames with specific properties.   It  also leads to fast  computational procedures
 and, at the same time, better de-noising precision since it exploits   both the sparsity and
the correlation structure  of the frame. Subsequently, we explore the case of a tight frame
and show how the techniques suggested for general frames are naturally simplified and speeded up in this situation.

Finally, we consider the case when a tight frame is formed as  a collection of orthonormal bases.
In this situation, hypothetically, one can obtain estimators for each of the orthonormal bases
separately and then combine them with  weights which sum up to unity.  Normally, in engineering practice,
the estimators are just combined with equal weights as it is done, for example,  in cycle spinning.
However, one can use different weights with the objective of obtaining an estimator
with better risk properties.   The  set of methods which exploit this idea is called {\it aggregation} and
was studied extensively by statistics community in the last decade (see,  for instance,
 Bunea   and  Nobel  (2008), Bunea,   Tsybakov and Wegkamp (2007),
Gribonval  (2003), Guleryuz (2007) , Juditsky  and Nemirovski  (2000),
 Juditsky,  Rigollet,    and Tsybakov  (2008), Leung  and  Barron  (2006),
Wegkamp  (2003) and  Yang (2001)).

Nevertheless, aggregation techniques have various limitations which make them
unsuitable for engineering practice. The existing techniques treat
estimators as constant (the risk is conditioned on those estimators) or require
sequential constructions of regression estimators and, in both cases,  lead to
expensive computational procedures. In particular, the algorithm of
Bunea,  Tsybakov and Wegkamp (2007) involves high-dimensional optimization
which is impossible to carry out in real-time computations.
Leung and Barron (2006), on the other hand, treat each of the regression estimators
as variable and work out an oracle expression for the risk which allows
them to offer  an explicit choice of weights. However, due to the fact that the estimators
are combined at the final stage, the authors cannot take full advantage of their
approach and are able to  combine   only one type of estimators, the least  squares estimators, in particular,
the least  squares estimators based on one basis function each.

In what follows, we take a more general and flexible approach to the aggregation problem.
We study the situation when both the risk and the estimators are variable
and they are combined before reconstruction, at the stage of frame coefficients.
In particular, we assume that a tight frame is constructed as  a collection of orthonormal bases.
The  frame coefficients   are subsequently de-noised and,  finally, the function is
re-constructed using variable weights for each   of the bases.
Using results of the earlier parts of the paper, we derive an oracle expression
for the risk which is not conditioned on a particular estimation strategy
and can take into account any explicit de-noising technique.  Moreover,
unlike in Leung and Barron (2006) and other aggregation papers, we derive an expression which
contains unknown weights in explicit form, making it easier to carry out necessary optimization.
Furthermore, our approach   allows one to explore both  the situation
of data independent weights  (or fixed estimatorsand) data dependent weights.
In the former case, we validate one of the main reasons for popularity of frames in engineering.
Indeed, we show that, if the frame is constructed as a combination of orthonormal bases,
then the risk of any frame estimator obtained  as a linear combination of the
estimators in each basis is smaller than the linear combination of the risks.
\\

The rest of the paper is organized as follows.
Section \ref{sec:OracleGeneralFrames} presents oracle expressions for the mean squared risks
of the diagonal shrinkage and thresholding estimators  in the case of
general or tight  frames.  Section \ref{sec:SURE_rules} provides SURE rules  for  those estimators.
Results obtained in Sections \ref{sec:OracleGeneralFrames} and \ref{sec:SURE_rules} are used in
Section \ref{sec:optimal} for designing optimal thresholding or shrinkage algorithms. Section
\ref{sec:ortho_collection} treats the case when the frame is constructed as a collection of orthonormal bases,
in particular, it gives recommendation how the estimators can be aggregated at the frame coefficient stage,
before reconstruction.  Section \ref{sec:simul} studies performances of the methodologies developed in the paper via numerical
simulations carried out on  test and real signals.  Section \ref{sec:discuss} concludes the paper with the discussion.
Finally Section \ref{sec:append} contains the proofs of the statements presented in the paper.

\section{Oracle expression for the risk for general or tight  frames}
\label{sec:OracleGeneralFrames}
\setcounter{equation}{0}

A collection of functions $\lfi w_i \rfi$ form a {\it frame }
in a separable Hilbert space $H$  if there exist two positive {\it frame bounds}
$C_l$ and $C_u>0$ such that, for any $f \in H$,
\be \label{frame_def1}
C_l \| f\|^2 \leq \sum_{i} |(f, w_i)|^2 \leq C_u \| f\|^2.
\ee
As particular cases of  frames one can
list Gabor frames, in which set $\lfi w_i \rfi$ comprises translated and modulated
versions of the same function, short time (or windowed) Fourier transform and wavelet frames.

In the space of discrete signals of length $n$, one usually considers $N$  vectors $w_i \in    C^n$, $i =1, \cdots, N$,
which together form matrix  $W \in  C^{N \times n}$ In these notations, \fr{frame_def1}   implies that $W$ is a matrix of a
{\it frame operator} if for any $f \in L_2 (R^n)$ one has
\be \label{frame_def2}
C_l \| f\|^2 \leq  f^* W^* W f \leq C_u \| f\|^2
\ee
where $W^*$ is a   transpose conjugate of $W$. The latter guarantees that eigenvalues of matrix $V = W^* W$
are bounded above and below and, therefore, $V$ is invertible.

If frame bounds are equal to each other, $C_l = C_u = \alpha$, then the frame is called {\it tight}
and $\alpha$ is referred to as a {\it frame constant}.
In the case of a tight frame the generalized Parseval's identity  holds and
$W^* W$ is proportional to the identity matrix.  In what follows, we shall assume that if the frame is tight, then
$W^* W = \af I_n$.  However, a tight frame can also be normalized so that $\alpha=1$,
 as it is done for the Gabor frame which is used for simulations in Section \ref{sec:simul}.

Consider a problem of recovering vector $f \in R^n$ from its noisy observation
\be \label{main_eq}
x = f + \delta,\ \ \delta \sim N(0, \sig^2 I_n).
\ee
Applying frame transform $W$ to both sides of equation \fr{main_eq},
obtain
\be \label{frame_eq}
y = \theta + \eps,\ \   \eps \sim N(0, \sig^2 U)
\ee
where $y = Wx$, $\theta = W f$, $\eps = W \delta$ and $U= W W^* \in C^{N \times N}$ .
The goal of the analysis is to reduce   noise in the vector of frame coefficients $y$
by shrinking or thresholding its components, thus, obtaining vector $\hat{\theta}$
and, subsequently, to estimate $f$ by
\be \label{est_of_f}
\hat{f} = V^{-1} W^* \hat{\theta} =  W^{+} \hat{\theta},
\ee
where $W^{+} = (W^* W)^{-1} W^*$ is the Moore–Penrose  inverse of matrix $W$.

We assume that the vector of frame coefficients $\theta$
is estimated by $\hat{\theta} = \Gamma y$ where $\Gamma = \diag(\gamma_1, \cdots, \gamma_N)$
is a fixed diagonal matrix in   $[0,1]^{N \times N}$.
The next statement provides an oracle expression for the risk  of this estimator.


\begin{theorem} \label{th:oracle}
If $\hat{\theta} = \Ga y$ where $\Ga$ is a fixed diagonal matrix, then
\be \label{lin_shrink_explicit}
\EE \| \hat{f} - f \|^2 =  \Tr[U^{-} (I_N - \Ga) \theta \theta^* (I_N - \Ga) + \sig^2 \Ga U \Ga U^{-}].
\ee
If the frame is tight, the previous expression takes the form
\be \label{lin_shrink_explicit_tight}
\EE \| \hat{f} - f \|^2 =  \alpha^{-2}\ \Tr[U (I_N - \Ga) \theta \theta^* (I_N - \Ga) + \sig^2 \Ga U \Ga U ].
\ee
Here, $U= W W^*$ and $U^{-} = (W^{+})^* W^{+}$.
\end{theorem}

Proofs of this and later statements  are given in Section \ref{sec:append}.
\\

Note that expressions \fr{lin_shrink_explicit} and \fr{lin_shrink_explicit_tight} require  simple minimization of quadratic forms
due to the following identity
\be     \label{oracle}
\operatornamewithlimits{argmin}_{\Ga=\diag(\gamma)} \left\{ \Tr[U^- (I_N - \Ga) \theta \theta^* (I_N - \Ga)
+ \sig^2 \Ga U \Ga U^- ] \right\}= \operatornamewithlimits{argmin}_{\gamma} \left\{ \gamma^* \tilde{A} \gamma -2  \gamma^* \tilde{b} \right\}
\ee
where $\tilde{A}=(\theta \theta^*) \circ U^-+\sigma^2 (U \circ U^-)$, $\tilde{b}=((\theta \theta^*) \circ U^-) e_N$,
$e_N$ is the vertical vector with all components equal to one and  $\circ$ denotes the Hadamard (element-wise) matrix product.
According to identity (\ref{oracle}), the optimal gain vector $\gamma=\diag(\Gamma)$   can be presented as
$$
\gamma=\left( \theta \theta^* \circ U^-+\sigma^2 U \circ U^- \right)^{-1} \left(\theta \theta^* \circ U^- \right) e_N
$$
and, in the case of tight frame,  it takes the form
\be \label{Wiener_U}
\gamma=\left( \theta \theta^* \circ U+\sigma^2 U \circ U \right)^{-1} \left(\theta \theta^* \circ U \right) e_N.
\ee
It is worth noting that, by construction, the weights $\gamma_i$
in the best  linear diagonal estimator are functions not only of $\theta_i$
but also of other coefficients in its neighborhood. In this sense,
the best  linear diagonal estimator is no longer diagonal  and
represents  an overlapping block  shrinkage procedure where the length of
the block is automatically determined by the correlations induced by the frame operator.
Moreover, observe that  matrix $\tilde{A}$ is invertible since the Hadamard product of
two positive-definite matrices is positive-definite. Moreover, matrix $U^-$ usually has
a block structure, so that the inversion of $A$ could be carried out  by fast algorithms specifically designed for this case.

According to Theorem~\ref{th:oracle},  for hard thresholding  one needs to  minimize   risk
\fr{lin_shrink_explicit} or \fr{lin_shrink_explicit_tight} over the set of arbitrary diagonal matrices
 with zero or unit values. Observe that in the case  of an orthonormal basis,
the oracle \fr{lin_shrink_explicit_tight} takes a familiar form
\beqns
\EE \| \hat{f} - f \|^2_{hard}  & = & \sum_{i=1}^n \lkv \theta_i^2 \II(\gamma_i=0)^2+ \sigma^2  \II(\gamma_i=1)\rkv
\eeqns
and motivates one to keep larger coefficients and discard smaller ones irrespective of the particular value of matrix $W$.
The situation changes when matrix $W$ ceases to be unitary.
Indeed, Theorem~\ref{th:oracle} implies that the choice of coefficients to  ``keep'' or ``kill'' depends
not only on their values but also on the entries of matrix $U$.

\section{SURE rules  for general or tight  frames}
\label{sec:SURE_rules}
\setcounter{equation}{0}

The advantage of the oracle expressions is that they allow to construct unbiased estimators
for the risk. Indeed,   matrix $\Theta = \theta \theta^*$ can be written as $\Theta = \EE (y y^*) - \sig^2 U$
and estimated by $\widehat{\Theta} = y y^* - \sig^2 U$. The latter leads to the following unbiased estimator for the risk:

\begin{corollary} \label{cor:lin_unbiased_risk}
If $\hat{\theta} = \Ga y$ where $\Ga$ is a fixed diagonal matrix, then
\be \label{expr_with_Delta}
\EE \| \hat{f} - f \|^2 = \sig^2 n + \EE  \Delta
\ee
where
\be \label{lin_shrink}
\Delta =   y^* (I_N - \Ga)  U^{-} (I_N - \Ga) y  - 2 \sig^2 \Tr[U^{-} U\, (I_N - \Ga) ].
\ee
In particular, if $\Gamma$ induces a hard thresholding rule, i.e. $\gamma_i=1$ or 0, then
\be \label{delta_hard}
\Delta = \sum_{i,j=1}^N \lkv y_i y_j U^{-}_{ij} - 2 \sigma^2 (U^{-} U)_{ii} \II(i=j) \rkv
\II (\gamma_i=0) \II (\gamma_j=0).
\ee
\end{corollary}

Since matrix $\Theta$ is non-negative definite,
all its diagonal elements should be non-negative which leads to the relations
$$
\hat{\Theta}_{ii} = y_i^2 - \sig^2 U_{ii} \geq 0.
$$
These inequalities themselves enforce  hard thresholds $\sig \sqrt{U_{ii}}$ on the values of $y_i$.
The oracle expression   \fr{lin_shrink_explicit_tight} allows for further reduction of the risk.

The oracles  \fr{lin_shrink_explicit} and  \fr{lin_shrink_explicit_tight}, though, are of  limited
value since they do  not allow one to access risk of more sophisticates
rules where matrix $\Gamma$ itself depends on $y$. In this case, one can write $\hat{\theta}$ as
\be \label{gen_est_theta}
\hat{\theta} = y + g(y).
\ee
Then, using modification of SURE, one obtains the following result:

\begin{theorem} \label{th:risk_genframes}
Let the data follow model \fr{main_eq} and $y$ be of the form \fr{frame_eq}.
Let $\hat{f}$ be given by formula \fr{est_of_f} with $\hat{\theta}$  of the form
\fr{gen_est_theta} where $g(y): R^N \rightarrow R^N$ is a continuous and piecewise differentiable column vector function.
Let $Z= \nabla_y g^* (y)$   be an $N \times N$-dimensional matrix with components
\be \label{matrix_Z}
Z_{ij} = \frac{\partial}{\partial y_i} \lkv g_j (y) \rkv.
\ee
Then,   the mean quadratic risk  is given by expression  \fr{expr_with_Delta}
with
\be \label{risk_gen_frames}
\Delta =   g^* (y) U^{-} g(y)  + 2 \sig^2 \Tr[U^{-} U\,  Z].
\ee
In the frame is tight,
then $U^{-} = \alpha^{-2} U$  and
\be \label{risk_tight_frames}
\Delta =   \alpha^{-2}\  g^* (y) U  g(y)  + 2 \sig^2 \alpha^{-1}\, \Tr[U \,  Z] .
\ee
\end{theorem}

Note that Theorem \ref{th:risk_genframes} allows one to obtain explicit expressions for various type of thresholding or shrinkage
procedures, as well as to construct unbiased estimators of the risk of those procedures.
If one uses linear shrinkage $\Gamma$, then  $g(y) = (\Ga - I_N) y$ and $Z= \Ga - I_N$, so that
Theorem \ref{th:risk_genframes} recovers expression \fr{lin_shrink} for $\Delta$.

In the case of soft thresholding with variable threshold $t_i$, one has
$ \hat{\theta}_i = (y_i - \sign(y_i) t_i) \II(|y_i|>t_i)$, so that
$g_i(y)$ is of the form
\be \label{soft_thresh}
g_i (y) = - \sign(y_i) \min(|y_i|, t_i),\ \ i=1, \cdots, N.
\ee
Hence,    $Z$ is a diagonal matrix with elements
\be \label{matrixZ}
Z_{ii} = -  \II(|y_i| < t_i)
\ee
and the following corollary is valid.

\begin{corollary} \label{cor:soft_thresh}
If $g(y)$ is defined by \fr{soft_thresh}, then the risk
is of the form \fr{expr_with_Delta} with
\be \label{Del_soft}
\Delta =  \sum_{i,j =1}^N \lkv   \sign (y_i y_j) \min(|y_i|, t_i) \min(|y_j|, t_j)  U^{-}_{ij} - 2 \sig^2 (U^{-} U)_{ii}\, \II(i=j)
\II(|y_i| < t_i)   \rkv.
\ee
If the frame is tight, the previous expression simplifies to
\be \label{Del_soft_tight}
\Delta =  \sum_{i,j =1}^N \lkv \alpha^{-2} \sign (y_i y_j) \min(|y_i|, t_i) \min(|y_j|, t_j) U_{ij}
- 2 \sig^2 \alpha^{-1}  U_{ii} \II(i=j)\ \II(|y_i| < t_i)   \rkv.
\ee
\end{corollary}

It is easy to check that, in the case of an orthonormal basis, familiar expressions
for the risks  can be easily recovered from formula \fr{Del_soft_tight}.
Indeed, setting $n=N$, $\alpha =1$ and  $U=I_n$, as before,
one obtains:
\beqns
\EE \| \hat{f} - f \|^2_{soft}  & = & \sig^2 n +  \EE  \lkv \sum_{i=1}^n  \min(y_i^2, t_i^2) -
2 \sig^2  \sum_{i=1}^n \II (|y_i| < t_i) \rkv.
\eeqns

\section {Designing optimal thresholding or shrinkage algorithms}
\label{sec:optimal}
\setcounter{equation}{0}

One can use expressions derived in the previous section to
design   an optimal shrinkage or thresholding strategy.
In what follows, just to be specific, we consider the case of linear shrinkage and
hard thresholding only. Other shrinkage or thresholding techniques including
soft thresholding can be analyzed in a similar manner.  Note that since matrices
$U$, $U^{-}$ and $U^{-} U$ are   not data-dependent, they may be calculated in advance
and, thus, the main computational complexity lies in solving the resulting optimization problems.

\subsection{Linear shrinkage}

Recall that the risk is of the form \fr{expr_with_Delta} where
$\Delta = \Tr[U^{-} (I_N - \Ga) y y^*  (I_N - \Ga)  + 2 \sig^2 U^{-} U \Ga
- 2 \sig^2  U^{-} U].$
Since the last term of the expression for $\Delta$ is independent of $\Ga$,
one needs to minimize
$$
F(\Ga) = \Tr[U^{-} (I_N - \Ga) y y^* (I_N - \Ga)  + 2 \sig^2 U^{-} U \Ga].
$$
Direct calculations show that this minimization takes the form of a quadratic programming problem.  Indeed,
if one define matrix $A = (y y^*) \circ U^{-}$ and vectors  $\ga = \diag(\Ga)$ and $b=(A-\sigma^2 U \circ U^{-}) e_N$,
then
\be \label{opt_problem}
F(\Ga) \equiv  F(\ga) = 
\ga^* A \ga - 2  \ga^* b,
\ee
and the optimal  $\ga \in  [0,1]^N$  which minimizes $F(\ga)$ takes the form
\be \label{sol_opt_problem}
\ga =((y y^*) \circ U^{-})^{-1} \left((y y^*) \circ U^{-} - \sigma^2 U \circ U^{-} \right) e_N.
\ee
Since matrices $U$, $U^{-}$ and $y y^*$ are nonnegative definite and Hermitian,
 matrix $A$ is also  nonnegative definite and Hermitian, and, thus,  the quadratic programming problem is convex.
Furthermore, note that matrix $A$  and vector $b$ are sparse.
For example, in the case of a tight frame, expressions for $A$ and $b$ take the forms
$A = \alpha^{-2} (y y^*) \circ U$ and $=(A-\sigma^2 U \circ U) e_N$.
Since the majority of entries of matrix $U$ are equal to zero, respective entries of matrix $A$
also vanish.


The optimization problem \fr{opt_problem} can also be modified by adding a penalty term $\pen(\ga)$  to  $F(\ga)$.
In particular, one can use  a quadratic penalty term $\gamma^* P \gamma$  with the positive definite matrix $P$
or   an $\ell_p$ penalty   of the  form
$\pen(\ga) = \beta \|\ga\|_{\ell_p}$ where $\|\cdot\|_{\ell_p}$ is a vector norm in ${\ell_p}$
space, which induces sparsity whenever $0 \leq p \leq 1$.

\subsection{Hard  thresholding }

In the case of hard thresholding, the SURE is of the form \fr{expr_with_Delta} with $\Delta$
given by expression \fr{delta_hard}. In order to minimize expression for $\Delta$ in
 \fr{delta_hard}, introduce matrix $H$ with components
$$
H_{ij} = \left\{
\begin{array}{ll}
y_i y_j U^{-}_{ij} & \mbox{if}\ \ \ i \neq j,\\
y_i^2 U^{-}_{ii} - 2 \sig^2  (U^{-} U)_{ii} & \mbox{if}\ \ \ i = j
\end{array} \right.
$$
Consider a set of indices $J$ such that $j \in J$ if $\gamma_j =0$ and $j \not\in J$ otherwise.
Then $\Delta$ can be re-written as
$$
\Delta = \sum_{i,j \in J} H_{ij}
$$
and the goal is to find a set  of indices $J$ such that the sum of respective row and column elements
of matrix $H$ is minimal.
This minimizations can be accomplished by a kind of a   greedy algorithm
which can be carried out as follows.
\\

\noindent
{\bf Greedy algorithm}

1. Since diagonal values of matrix $H$ are counted once while all other elements are counted twice,
introduce modified matrix $\tilde{H}$ with elements
$$
\tilde{H}_{ij} = \left\{
\begin{array}{ll}
H_{ij},  & \mbox{if}\ \ \ i \neq j,\\
H_{ij}/2,  & \mbox{if}\ \ \ i = j
\end{array} \right.
$$
Set $J= \left\{ 1, \cdots, N \right\}$.

2. Find  a column $l$ of $\tilde{H}$ with the maximum sum of elements.

3. If the sum of elements of column $l$ is positive, then eliminate column $l$ and row $l$
from $\tilde{H}$ and index $l$ from set $J$, and RETURN TO STEP 2.
If the sum of elements of column $l$ is zero or negative,  then FINISH.

4. Set $\gamma_j=0$ if $j \in J$ and $\gamma_j=1$ if $j \not\in J$.

\section {Frame constructed as a collection of orthonormal bases}
\label{sec:ortho_collection}
\setcounter{equation}{0}

Consider the case when a frame is constructed as a collection of
$m$ orthonormal bases. In this case, $N = nm$ and matrix $W$ has a block structure
with $m$ vertical blocks $W^{(i)} \in C^{n \times n}$,   $i=1, \cdots, m$, such that
$W^{(i)} (W^{(i)})^* = (W^{(i)})^* W^{(i)} = I_n.$
Denote $U^{(i,j)} = W^{(i)} (W^{(j)})^*$. Then, matrix $U$ is a block matrix with blocks
$U^{(i,j)}$, $i,j = 1, \cdots, m,$ and $U^{(i,i)} = I_n$.
It is easy to see that $W$ constitutes a tight frame with $\alpha = m$.

An interesting phenomenon for the frame of this type is that, since each of matrices $W^{(i)}$
allows complete reconstruction of $f$, one can combine those reconstructions with non-equal weights.
Let $\Lam$ be a block-diagonal matrix with   blocks $\Lam^{(i,i)} = \lam_i I_n$, $i=1, \cdots, m$,
where weights $\lam_i$ sum to unity:
\be \label{lam_condition}
\sum_{i=1}^m \lam_i =1.
\ee
Note that, under condition \fr{lam_condition}, one has
$$
W^* \Lam W = \sum_{i=1}^m  (W^{(i)})^* \Lam_i W^{(i)} =
\sum_{i=1}^m \lam_i (W^{(i)})^*   W^{(i)} = I_n.
$$
Therefore, if $\theta = W f$, then $f$ can be reconstructed as
$f = W^* \Lam \theta$ and estimated by
\be  \label{f_block_est}
\hat{f} = W^* \Lam \hat{\theta}.
\ee
Usually, in the current engineering practices, weights are chosen to be equal (as in, e.g.,
cycle spinning), however, this choice does not allow one to reduce the total risk
by assigning a smaller weight to an estimator with a higher risk.

It is easy to see that the problem of choosing weights in this set up is ultimately related
with aggregation problem studied in, for instance,
 Bunea   and  Nobel  (2008), Bunea,   Tsybakov and Wegkamp (2007),
Gribonval  (2003), Guleryuz (2007) , Juditsky  and Nemirovski  (2000),
 Juditsky,  Rigollet,    and Tsybakov  (2008), Leung  and  Barron  (2006),
Wegkamp  (2003) and  Yang (2001) among others.
Indeed, note that one can consider an estimator of $f$ of the form
$\hat{f}^{(i)} = (W^{(i)})^* \hat{\theta}^{(i)}$ where
$\hat{\theta}^{(i)} = y^{(i)} + g^{(i)} (y^{(i)})$ is an estimator of $\theta^{(i)} = W^{(i)} f$,
the coefficients of representation of $f$ in the $i$-th basis. Then,
estimator $\hat{f}$ in \fr{f_block_est} can be re-written as
\be \label{aggr_reconstr}
\hat{f} = \sum_{i=1}^m \lam_i \hat{f}^{(i)}.
\ee
The difference between our approach and aggregation techniques, however, is that we carry out
aggregation at the level of frame coefficients, not estimators of $f$ themselves.
This will allow us to avoid, if desired, both conditioning on estimation technique
(``constant estimators'') and treating weights as data-independent.

Nevertheless, we shall start with the case of data-independent weights and then
investigate a more elaborate case where weights are data-dependent.

\subsection{Data independent weights: better than the best basis}

If the weights are data independent, then direct calculations show that
\be \label{risk_exp1_di_weights}
\EE \| \hat{f} - f \|^2 = \sum_{i,j=1}^m \lam_i \lam_j  \rho_{ij}\ \ \ \mbox{with}
\ \ \ \rho_{ij} = \EE [(\hat{\theta}^{(i)} - \theta^{(i)})^* U^{(i,j)} (\hat{\theta}^{(j)} - \theta^{(j)})].
\ee
The statement below shows that the error of $\hat{f}$ is always smaller than the weighted sum of the errors
of estimators  $\hat{f}^{(i)}$.

\begin{theorem} \label{th:const_weights}
If $\hat{f}$ is defined in \fr{f_block_est} and weights $\lam_i$ are data independent and
satisfy condition \fr{lam_condition}, then
\be \label{risk_rel_di_weights}
\EE \| \hat{f} - f \|^2 = \sum_{i=1}^m \lam_i \EE \lkv \|\hat{f}^{(i)} - f \|^2 - \|\hat{f}^{(i)} - \hat{f}\|^2 \rkv,
\ee
so that if, furthermore, the weights are  nonnegative,
$$
\EE \| \hat{f} - f \|^2 \leq \sum_{i=1}^m \lam_i \EE \|\hat{f}^{(i)} - f \|^2.
$$
\end{theorem}

Theorem \ref{th:const_weights} does not allow one to choose optimal weights, since
weights enter expression \fr{risk_rel_di_weights} implicitly in the form of
 $\hat{f}$. In order to evaluate expression for the risk in the case of constant weights, one
needs to combine expression  \fr{risk_exp1_di_weights} and  formula \fr{main_formula}
with $g^{(i)}(y^{(i)})$, $U^{(i,j)}$ and $Z^{(j,j)}$ instead of $g (y)$, $U$ and $Z$, respectively:
$$
\EE \| \hat{f} - f \|^2   =   \sum_{i,j=1}^m \lam_i \lam_j \left\{ \sig^2 \Tr[U^{(i,j)} U^{(j,i)}] +
\EE [(g^{(j)})^*  U^{(j,i)} g^{(i)}] + 2 \sig^2 \Tr[U^{(i,j)} Z^{(j,j)} U^{(j,i)}] \right\},
$$
where for the sake of brevity, we denoted $g^{(i)}(y^{(i)}) = g^{(i)}$.
Taking into account that $U^{(i,i)} = I_n$ and $U^{(i,j)} U^{(j,i)}=I_n$, one arrives at the risk
of the form  \fr{expr_with_Delta} with
\be \label{Delta_di_weights}
\Delta =   \sum_{i,j=1}^m \lam_i \lam_j   (g^{(i)})^* U^{(i,j)} g^{(j)}
+ 2 \sig^2  \sum_{i=1}^m \lam_i  \Tr [ Z^{(i,i)}].
\ee
The above expression contains weights in explicit form and allows to choose optimal weights for any
kind of a shrinkage or thresholding technique. Observe that the choice of a ``best'' basis corresponds
to one of the coefficients $\lam_i$ being one and the others being zero. This would be an optimal choice
if matrix $\rho$ with entries $\rho_{ij}$ defined in \fr{risk_exp1_di_weights} were diagonal.
However, since this is not the case, the choice of only one estimator versus a mixture may not be the best strategy,
both, from the point of view of risk and even sparsity (see, e.g., Elad  and  Yavneh  (2009)).

\subsection{Data dependent weights}

In order to study the case of data--dependent weights, recall that $y \in R^N$ is a  vector with $m$ block-components
$y^{(i)} = \theta^{(i)} + \eps^{(i)},$ where $\eps^{(i)} \sim N(0, \sig^2 I_n)$
and $\theta^{(i)}$ is estimated by $\hat{\theta}^{(i)}  = y^{(i)} + g^{(i)} (y^{(i)})$.
Introduce data dependent weights $\lam_i(y)$ such that relation \fr{lam_condition}
is valid for any value of $y$.

Here, we ought to point out two essential features of our choice of weights. First,
we explicitly choose weights depending on frame coefficients $y$ rather than raw data $x$.
Second, weights for each basis depend on all frame coefficients.
Re-writing \fr{f_block_est}, we obtain
$$
\hat{f} = \sum_{i=1}^m \lam_i(y)\, (W^{(i)})^*\, \hat{\theta}^{(i)}
= \sum_{i=1}^m (W^{(i)})^* \left\{ \lam_i(y) [ y^{(i)} + g^{(i)} (y^{(i)})] \right\}.
$$
Note that $\hat{f}$ in the last expression can be presented as  $\hat{f} = m^{-1} W^* \tilde{\theta}$,
the tight frame reconstruction of the estimator
$\tilde{\theta} = \tilde{\theta} (y) = y + \tilde{g}(y)$ of frame coefficients.
Here $\tilde{g}(y)$ is a block vector with blocks
\be \label{tildeg}
\tilde{g}^{(i)} (y) = m \lam_i(y) [y^{(i)} +  g^{(i)} (y^{(i)})] - y^{(i)}.
\ee
Hence, we can use expression \fr{risk_tight_frames} in Theorem \ref{th:risk_genframes}
with $\alpha =m$ and $\tilde{g}$ and $\tilde{Z}$ instead of $g$ and $Z$, respectively.

\begin{theorem} \label{th:risk_dd_weights}
If $\hat{f}$ is defined in \fr{f_block_est} and weights $\lam_i = \lam_i (y)$ are data--dependent and
satisfy condition \fr{lam_condition} for every $y$, then the risk is of the form  \fr{expr_with_Delta} with
\beqn
\Delta & = & \sum_{i,j=1}^m \lam_i(y) \lam_j(y)  (g^{(i)})^*   U^{(i,j)} g^{(j)}
  +    2 \sig^2 \sum_{i=1}^m \lam_i (y) \Tr(Z^{(i,i)}) + \Delta_0, \label{Delta_dd_weights} \\
\Delta_0 & = & 2 \sig^2 \sum_{i,j=1}^m (\hat{\theta}^{(i)})^*   U^{(i,j)} \lkv \nabla_{y^{(j)}} \lam_i (y) \rkv.
\nonumber
\eeqn
Here, same as before, $\hat{\theta}^{(i)} (y^{(i)}) = y^{(i)} + g^{(i)} (y^{(i)})$, and, for the sake of brevity, we
omitted $(y^{(i)})$ in the expressions $g^{(i)} (y^{(i)})$ and $\hat{\theta}^{(i)} (y^{(i)})$.
\end{theorem}

Straightforward comparison shows that the first two terms in   \fr{Delta_dd_weights} coincide
with the respective terms in \fr{Delta_di_weights} while the last term vanishes when the weights are data independent.
Expression  \fr{Delta_dd_weights} contains weights explicitly, so, hypothetically, it can be used for choosing
data dependent weights.

The difficulty with using formula  \fr{Delta_dd_weights}, however, lies in the fact that one would like to choose
weights depending not on frame coefficients $y$ but rather on the risk of the $i$-th estimator $\hat{\theta}^{(i)} (y^{(i)})$, or,
more precisely, on the Stein unbiased estimator of this risk.  For this reason, one needs to learn how to find partial derivatives
of the unbiased estimator of the risk,  which is accomplished by the following statement.

\begin{lemma} \label{lem:SURE_deriv}
Let the data follow model $y = \theta + \eps$ where $y, \theta, \eps \in R^n$ and
$\eps \sim N(0, \sig^2 I_n)$. Let $\hat{\theta}$ be an estimator of $\theta$ of the
form $\hat{\theta} (y) = y + g(y)$. Then for the Stein unbiased estimator
$$
r(y) = \sig^2 n +  g^*(y) g(y) + 2 \sig^2 \Tr \lkv \nabla_y g^* (y) \rkv
$$
of the risk $\EE \| \hat{\theta} (y) - \theta \|^2$ one has
\be \label{deriv_expr}
\nabla_y r(y) =  2 [ \nabla_y g^* (y) ] g(y) + 2 \sig^2 d(y)
\ee
where $d(y)$ is a column vector with components
\be  \label{d_components}
d_k(y) = \sum_{l=1}^n \frac{\partial^2 g_l (y)}{\partial y_l \partial y_k},\ \  k=1, \cdots, n.
\ee
\end{lemma}

Recall that in the case of fixed linear shrinkage $g_l(y) = (\Ga_l - 1)y_l$
and for soft thresholding $g_l(y) = - \sign(y_l) \, \min(y_l,t)$, where $t$ is the threshold,
one has $d(y) =0$. Therefore, in those two cases,
$\nabla_y r(y) =  2 [ \nabla_y g^* (y) ] g(y)$.

Expression \fr{Delta_dd_weights} can be potentially used in order to minimize $\Delta$
with respect to weights. However, this expression is too general to use.
Hence, following Leung  and  Barron  (2006), we consider weights in the exponential form.

\subsection{Weights in the exponential form}

Let the weights be of the  form
\be \label{expo_weights}
\lam_i (y) = \frac{\pi_i \exp \lkv - \beta \eta_i (y^{(i)}) \rkv}
{\sum_{l=1}^m  \pi_l \exp \lkv - \beta \eta_l (y^{(l)}) \rkv}
\ee
where $\pi_i \geq 0$, $i=1, \cdots, m$.
Presentation \fr{expo_weights} guarantees that the weights $\lam_i (y)$ sum to unity.
Usually, the most intuitive choice is $\eta_i = r_i (y^{(i)})$, the SURE
of the $i$th estimator $\hat{\theta}^{(i)} (y^{(i)})$.

The following corollary of Theorem \ref{th:risk_dd_weights} provides an explicit expression for the unbiased
estimator of the risk for the weights in the form \fr{expo_weights}.

\begin{corollary} \label{cor:Delta0_expo_weights}
If $\hat{f}$ is defined in \fr{f_block_est} and weights  are in the form \fr{expo_weights},
then the risk is of the form  \fr{expr_with_Delta} with $\Delta$ given by formula \fr{Delta_dd_weights}
and
\beqn
&&\Delta_0   =    2 \sig^2 \beta \sum_{i=1}^m \lam_i (y) \lkv \nabla_{y^{(i)}} \eta_i (y^{(i)}) \rkv^*
\lkr W^{(i)} \hat{f}  - \hat{\theta}^{(i)}  \rkr \label{DD1_form1}\\
&& =   2 \sig^2 \beta \lfi \sum_{i,j=1}^m \lam_i(y) \lam_j(y) \lkv \nabla_{y^{(j)}} \eta_j (y^{(j)}) \rkv^* U^{(j,i)} \hat{\theta}^{(i)}
- \sum_{i=1}^m \lam_i (y)\, \lkv \nabla_{y^{(i)}} \eta_i (y^{(i)}) \rkv^* \hat{\theta}^{(i)} \rfi .\ \ \ \  \label{DD1_form2}
\eeqn
Here, as before, $\hat{\theta}^{(i)}   = y^{(i)} + g^{(i)}$.
\end{corollary}

Note that representation \fr{DD1_form1} of the risk is more compact but does not contain the weights explicitly,
while formula \fr{DD1_form2} is more convenient if one wants to minimize the risk with respect to
$\pi_i$, $i=1, \cdots, m,$ or $\beta$.

If $\eta_i (y^{(i)}) = r_i(y^{(i)})$, $i=1, \cdots, m$, where $r_i(y^{(i)})$ is the unbiased estimator of the
risk
$$
r_i(y^{(i)}) = \sig^2 n +  [g^{(i)}(y^{(i)})]^* g^{(i)}(y^{(i)}) + 2 \sig^2 \Tr \lkv \nabla_{y^{(i)}} [g^{(i)}(y^{(i)})]^* \rkv,
$$
of the estimator $\hat{f}^{(i)}$, then, by Lemma \ref{lem:SURE_deriv}, one has
$$
\nabla_{y^{(i)}} r_i(y^{(i)}) =  2 \lkv \nabla_{y^{(i)}} [g^{(i)}(y^{(i)})]^* \rkv g^{(i)} (y^{(i)}) + 2 \sig^2 d_i(y^{(i)})
$$
where $d_i(y^{(i)})$ is a column vector with components
$$
d_{ik} = \sum_{l=1}^n \frac{\partial^2 g^{(i)}_l (y^{(i)})}{\partial y^{(i)}_l \partial y^{(i)}_k},\ \  k=1, \cdots, n.
$$
In particular, if one uses linear shrinkage or thresholding, soft or hard, then $d_i(y^{(i)})=0.$

\section {Simulation Study}
\label{sec:simul}
\setcounter{equation}{0}

\begin{figure}
\centering
\includegraphics[width=4in,height=4in]{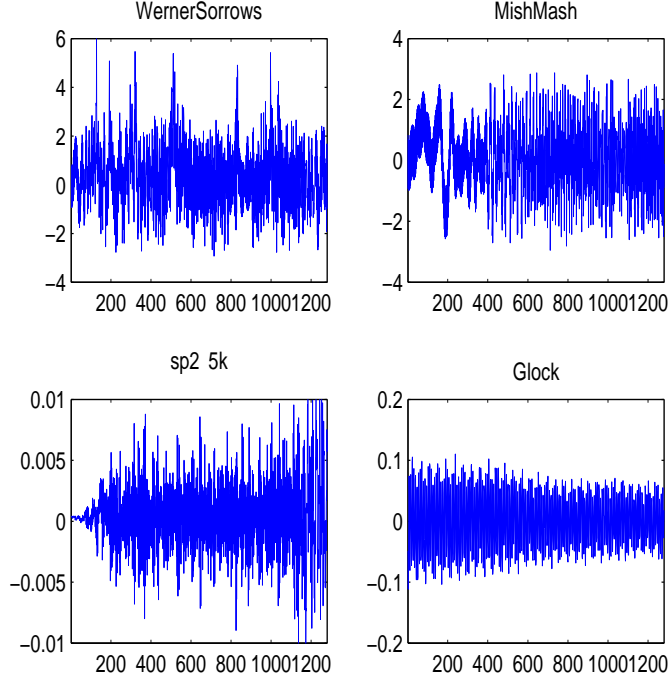}
\caption{ Normalized test signals of length 1280. } \label{true_signal}
\end{figure}

In this section, we carry out some numerical experiments to study the finite sample performances of the proposed estimators.
It is well known that the choice of a frame is linked to the underlying function $f$ to be de-noised.
The advantage of using  a frame  compared to an orthogonal   basis is that it can provide an efficient
representation of a broad class of signals as well as better adaptivity for their parsimonious representation.
In our simulation study,  we use the classical Gabor frame with Hamming window. This  is a tight frame which  is
particularly suitable for representation of fast oscillating signals such as audio signals.
For that reason, we consider two fast oscillating standard test signals,  {\it WernerSorrows} and {\it Mishmash},
reproducible by MakeSignal of the toolbox Wavelab, and two pieces of real audio signals {\it sp2-5k.wav} and
{\it Glock.wav}. The test signals listed above are displayed  in Figure~\ref{true_signal}.

The objective of this simulation study is to illustrate the gain in de-noising precision
obtained by taking into the account the frame structure  rather than be an exhaustive study of signal de-noising by frames.
Results of all comparisons are represented in terms of the means and the standard deviations of the $L_2$ errors.
In order to show the advantage attained by accounting for the frame structure, we compare the ideal best diagonal
estimator obtained minimizing the true risk in (\ref{Wiener_U}) versus the ideal best diagonal estimator obtained by minimizing
the true risk without taking into account the frame structure i.e., considering $U=I$.
We denote these two estimators $IDEAL_U$ and $IDEAL_I$, respectively. Note that estimators $IDEAL_U$ and $IDEAL_I$
are not available in practice, but their comparison can give  an idea of the {\bf best} possible gain obtained by taking into account
the frame structure. The empirical versions of these estimators are derived by substituting $\theta \theta^*$
with its unbiased estimator $y y^*-\sigma^2 U$ and  $y y^*-\sigma^2 I_N$, respectively. In the first case, we obtain
$$
\gamma=\left( y y^T \circ U \right)^{-1} \left((y y^T-\sigma^2  U) \circ U \right) e_N
$$
which coincides with solution  (\ref{sol_opt_problem}) in the case $U^-=U$,
while, in the second case, we obtain
 $$
\gamma=\left( y y^T \circ I_N \right)^{-1} \left((y y^T-\sigma^2 I_N ) \circ I_N \right) e_N
$$
which component-wise reduces to the well known empirical  Wiener filter  $\gamma_i=(y_i^2-\sigma^2)/y_i^2$.
In what follows, we refer to  these estimators as  $EMP_U$ and $EMP_I$, respectively.  Since matrices
$\left( y y^T \circ U \right)$ and especially $\left( y y^T \circ I_N \right)$ sometimes have high condition numbers,
in order to stabilize their inversion in our simulation study,  we add a quadratic penalization term
$\gamma^t P \gamma$ to the functional  with matrix $P=\zeta \ I_N$.

Results for $\zeta =10^{-4.5}$ are reported  in Table \ref{tab_res1} and are based on 100 simulational runs
with  signal-to-noise ratios (SNR)  1, 3 and 5, which represent, respectively,  severe,  moderate
and low  noise levels.   As it is standard in the statistical   literature, the  signal-to-noise ratio (SNR)
is defined here as the ratio of the standard deviations of the signal and the noise.
The empirical estimators  $EMP_U$ and $EMP_I$ approximate the corresponding ideal estimators $IDEAL_U$ and $IDEAL_I$
when the  noise level is low (SNR=5) and may be quite far from them when the  noise level is high   (SNR=1).
However, for all the test signals, the ideal gain (the difference between the first and the second columns)
and the empirical gain (the difference between the third and the fourth columns) obtained by accounting
for the frame structure is quite significant, especially, in the case of severe noise.

In a similar manner, we carry out comparisons between soft thresholding  procedures obtained with and without
consideration of the specific frame structure. In particular, we construct estimators $SOFT_U$ and  $SOFT_I$
which are obtained by the formula $\hat{\theta}_i=(y_i- \sign(y_i)t) \II(|y_i| > t)$
with the global threshold $t$ obtained by minimizing, respectively,
expression (\ref{Del_soft_tight}) when $t_i=t$  for all $i=1,\cdots, N$, and
\be
\operatornamewithlimits{argmin}_{t} \left\{  \sum_{i=1}^n  \min(y_i^2, t^2) -
2 \sig^2  \sum_{i=1}^n \II (|y_i| < t) \right\}.
\ee
which is the classical expression of the SURE reported in Donoho and Jonstone (1995).
 Similarly, we  compare estimators $VISU_U$ and $VISU_I$ obtained using hard thresholding procedure
$\hat{\theta}_i=y_i \II(|y_i| > t)$ where, in the first case, the expression for the universal
threshold is provided   in Haltmeier and Munk (2012)
$$
t=\sigma \sqrt{2 \log N} + \sigma \left(\frac{2 z-\log(\log N)-\log \pi }{2 \sqrt{2 \log N}}   \right)
$$
with $z=\pi/\sqrt{6}$,  and, in the second case, $t$ is the classical universal threshold for the
orthonormal bases  $t=\sigma \sqrt{2 \log N}$.

Results of comparisons are reported  in Table \ref{tab_res2}  and are based on 100 simulation  runs.
It is easy to notice that the gain obtained by taking frame structure into account is much more significant
in the case of SURE than for both  $VISU_U$ and $VISU_I$ universal thresholding procedures. This is due to the fact
that universal threshold is known to be  too large for de-noising applications, as it has been already noted
in statistical literature (see, e.g.,  Donoho and Johnstone (1995)). In fact, SURE-based soft thresholding procedures
outperform the universal thresholding procedures even if the former does not take the fame structure into account:
it follows from Table  \ref{tab_res2} that $SOFT_I$ has better precision than  $VISU_U$ for every test signal and
every noise level.

\vspace{1mm}

\begin{table}
\caption{Results obtained over 100 runs and with parameter choices,  $n=1280$ and 64-sampled Hamming window.}
\label{tab_res1}
\begin{center}\scriptsize
\begin{tabular}{ c | c c c c }
  &  $IDEAL_U$ &    $IDEAL_I$   &  $EMP_U$ &   $EMP_I$ \\
\hline
{ \sc WernerSorrows} &   &  &      &    \\
  SNR=1  &  0.1327 (0.0096)  &  0.2274 (0.0116) & 0.4964 (0.0240)  &    5.7420 (0.1847)     \\
  \\
 SNR=3  &   0.0284  (0.0019) &  0.0404  (0.0022) & 0.0777 (0.0034)  &   0.1343 (0.0049)    \\
 \\
 SNR=5  &  0.0126  (0.0006)&  0.0167 (0.0007 )&   0.0321  (0.0011 )&  0.0412 (0.0015) \\
  \hline
 {\sc MishMash}  &   &  &    &   \\
SNR=1 &  0.1026 (0.0106) & 0.1837 (0.0139) &  0.4881 (0.0254)  &    6.2411  ( 0.2290) \\
\\
SNR=3 &  0.0211 (0.0017)  & 0.0284  (0.0021) &  0.0752 (0.0036)  &  0.1113  (0.0044)\\
\\
SNR=5   &  0.0094 (0.0007)&   0.0122 (0.0008)&   0.0324  (0.0013)&  0.0286 (0.0015) \\
 \\
\hline
{\sc sp2-5k}  &   &  &      &    \\
SNR=1  & 0.1533(0.0112)   &  0.2474(0.0127)  &  0.5201 (0.0254)&   6.2648(0.1745) \\
 \\
 SNR=3  &  0.0363 (0.0022)  &0.0548  (0.0024) & 0.0849 (0.0039) &   0.1771 (0.0058)  \\
\\
SNR=5   &   0.0168  (0.0009)&  0.0244 (0.0011)&   0.0349   (0.0014)& 0.0614  (0.0022) \\
 \\
\hline
 {\sc Glock}  &   &  &    &  \\
  SNR=1  &   0.0845(0.0075)   &  0.1305 (0.0093)  &  0.4529 (0.0245) &  6.4889 (0.2079) \\
  \\
 SNR=3  &  0.0192 (0.0014)   &  0.0278  (0.0016)&   0.0737 (0.0037)& 0.1232 (0.0043) \\
 \\
 SNR=5   &  0.0089  (0.0006)&  0.0123  (0.0007)&  0.0322 (0.0013)&   0.0326 (0.0015)\\
 \\

 \end{tabular}
\end{center}
\end{table}

\begin{table}
\caption{Results obtained over 100 runs and with parameter choices,  $n=1280$ and 64-sampled Hamming window.}
\label{tab_res2}
\begin{center}\scriptsize
\begin{tabular}{ c | c c c c c c}
  &   $SOFT_U$ &  $SOFT_I$ & $VISU_U$ & $VISU_I$\\
\hline
{ \sc WernerSorrows} &   &  &      &   &   &   \\
  SNR=1  &        0.3748 (0.0188)&   0.8511  (0.0414)& 0.8987  (0.0199)&  0.9024 (0.0200)\\
  \\
 SNR=3  &      0.0763 ( 0.0041)&   0.1342   (0.0114) & 0.3748  (0.3748)&  0.3965 (0.3965)\\
 \\
 SNR=5  &   0.0327 (0.0016)&    0.0481 (0.0039) & 0.1230  (0.0041)&  0.1275 (0.0041)\\
  \hline
 {\sc MishMash}  &   &  &    & &   &  \\
SNR=1 &       0.3519   (0.0216)& 0.8970   (0.0695) &  0.9733  (0.0173)&  0.9756 (0.0158)\\
\\
SNR=3 &    0.0602  (0.0040)&  0.1063 (0.0095) &  0.2434 (0.0148)&   0.2573 (0.0160)\\
\\
SNR=5   &    0.0251  (0.0013)&  0.0414 (0.0035) &  0.0749 (0.0039)&   0.0786 (0.0039)\\
 \\
\hline
{\sc sp2-5k}  &   &  &      &   &  &    \\
SNR=1  &   0.3893 (0.0197)  & 0.8555(0.0693)  & 0.9934 (0.0120)&   0.9952 (0.0105)\\
 \\
 SNR=3  &   0.0917 (0.0047) &   0.1689  (0.0164) &  0.3881  (0.0143)&  0.4023 (0.0142)\\
\\
SNR=5   &     0.0457 (0.0026)&   0.0626 (0.0059) & 0.1740  (0.0052)&  0.1799 (0.0050) \\
 \\
\hline
 {\sc Glock}  &   &  &    & &   &  \\
  SNR=1  &    0.2853 (0.0186)&   0.5064 (0.0462) & 0.9181  (0.0453)&  0.9350 (0.0462)\\
  \\
 SNR=3  &   0.0516 (0.0029) &   0.0981 (0.0083) & 0.1591 (0.0076)&   0.1628 (0.0078) \\
 \\
 SNR=5   &    0.0228 (0.0012) & 0.0406 (0.0034) &  0.0898   (0.0026)& 0.0919 ( 0.0026)\\
 \\

 \end{tabular}
\end{center}
\end{table}

\begin{figure}
\centering
\includegraphics[width=4in,height=4in]{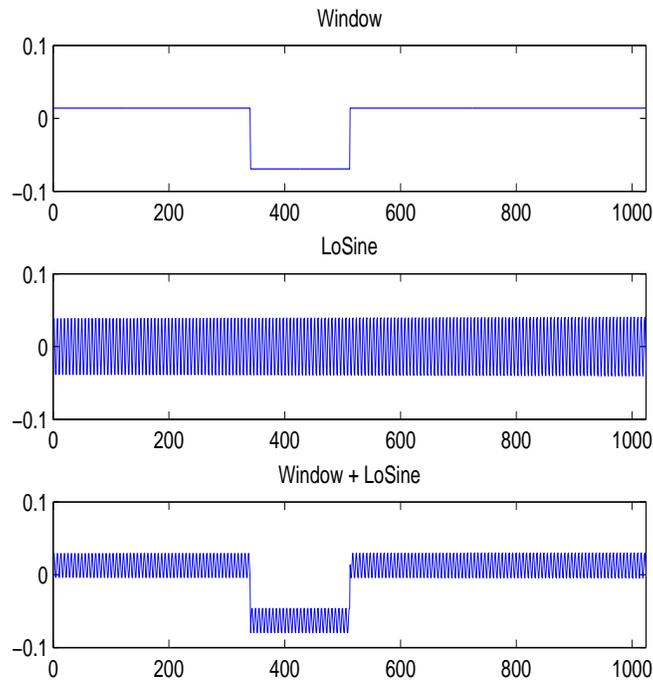}
\caption{ Normalized test signals of length 1024. } \label{window_signal}
\end{figure}

In order to  examine  the performance of the estimator proposed in Section 5, we study
the simple case of data independent weights. In particular, we consider two classical
orthonormal bases,  {\sc Cosine} and {\sc Haar}, and   three test functions, {\sc Window}, {\sc LoSine} and a combination of the two (see Figure~\ref{window_signal}). The  {\sc Window} and the {\sc LoSine} are classical test signals which are very well represented,
respectively,  by the {\sc Haar} and the  {\sc Cosine} bases.
We evaluate  estimator (\ref{aggr_reconstr}), where $\lambda$ is derived by minimizing expression
(\ref{Delta_di_weights}) and $\hat{f}^{(i)}$'s are obtained as soft thresholding estimators with the
universal data-independent threshold. The risks of the estimators  are presented in the forth column of Table~\ref{Tab3}
and  the mean values of the estimated weights $\lambda$ are displayed in the fifth column. Table~\ref{Tab3} also
reports  the risks of the single estimators (columns one and two)  and of the  average of the estimators  obtained with the weights
 $ \lambda_1 = \lambda_2 = 0.5$. Note that  the aggregation estimator  is always  better or at least as good as the best basis
estimator and it is always better then the  estimator obtained by simple average (i.e., by the default frame reconstruction).
Moreover, it is instructional to observe that the choice of weights  $\lambda_i$  supplied by criterion (\ref{Delta_di_weights})
follows an intuitive preference. Indeed, one would favor {\sc Cosine} basis for {\sc LoSine} signal, {\sc Haar} basis for {\sc Window}
signal as well as a balanced combination of the two bases for the sum of these two signals: computations confirm those intuitive assessments.


\begin{table}
\caption{Results obtained over 100 runs by {\sc Cosine} and {\sc Haar}  bases. Parameter choices are $n=2^{10}$ and $J=3$ for the Haar basis.}
\label{Tab3}
\begin{center}\scriptsize
\begin{tabular}{ c |c | c c c c c }
 function&  SNR    &   {\sc Cosine}  &  {\sc Haar}    & averaging  &aggregation & ($\lambda_1$ $\lambda_2$)\\

  \hline
{\sc Window} & 1 & 0.2291 &   0.1719   & 0.1719 &   0.1648 &  (0.2200    0.7800)\\
 & 3 &   0.0742  &  0.0214   & 0.0364  &  0.0214  &   (0.0105    0.9895)\\
 & 5 &   0.0444   & 0.0076   & 0.0182  &  0.0077  & (0.0045    0.9955) \\
 \hline
{\sc LoSine} & 1 &    0.1284 &   0.9940  &  0.4118  &  0.1284 &  (1.0000    0.0000)\\
 & 3 &    0.0427  &  0.6682   & 0.2221 &   0.0444   & (0.9836    0.0164)\\
 & 5 &   0.0259   & 0.3617  &  0.1177  &  0.0305    & (0.9181    0.0819)\\
   \hline
 {\sc Window + LoSine  } & 1 &     0.2644   & 0.3496   & 0.2673   & 0.2563 &   (0.7693    0.2307)\\
  & 3 &   0.0855   & 0.2011  &  0.1037 &   0.0827  &  (0.8650    0.1350)\\
  & 5 &  0.0509  &  0.1666   & 0.0720   & 0.0496 &   (0.8448    0.1552) \\
     \hline
\end{tabular}
\end{center}
\end{table}

\section {Discussion}
\label{sec:discuss}
\setcounter{equation}{0}

The present paper provides a  comprehensive  study of   de-noising
properties of frames and, in particular, tight frames, which constitute one of the
most popular tools in contemporary signal processing.  The objective of the  paper is to bridge the
existing gap between mathematical and statistical theories on one hand and engineering practice on the other
and explore how one can take advantage of a specific structure of a frame
in contrast to an arbitrary collection of vectors or an orthonormal basis.

For both the general and the tight frames, the paper presents
a set of practically implementable de-noising techniques which take frame induced correlation structures
into account.  These results are supplemented by an examination of the case when the frame is constructed as
 a collection of orthonormal bases. In particular, recommendations are given for aggregation of the
estimators at the stage of frame coefficients. The paper is concluded by a finite sample simulation study
which confirms that taking frame structure and frame induced correlations  into account indeed improves de-noising precision.

\section*{Acknowledgements}

Marianna Pensky was supported in part by National Science Foundation
(NSF), grant  DMS-1106564.

\section {Appendix}
\label{sec:append}
\setcounter{equation}{0}

\noindent
{\bf Proof of Theorem  \ref{th:oracle}. }
To verify expression \fr{lin_shrink_explicit}, note that
$$
\EE \| \hat{f} - f \|^2 =  \Tr[W^+ (\Gamma  y - \theta)(\Gamma  y - \theta)^* (W^+)^*] = \Delta_1 + \Delta_2
$$
where
\beqns
\Delta_1 &=& \Tr [W^+  \Gamma \EE (\epsilon \epsilon^*) \Gamma^* (W^+)^*] = \sig^2 \Tr [\Ga U \Ga U^{-}]
\eeqns
and
 \beqns
\Delta_2 &=& \Tr [W^+ (I_N - \Ga) \theta \theta^*  (I_N - \Ga) (W^+)^*] =
\Tr[U^{-} (I_N - \Ga) \theta \theta^* (I_N - \Ga)],
\eeqns
which completes the proof of \fr{lin_shrink_explicit}. To prove \fr{lin_shrink_explicit_tight}, note that
in the case of a tight frame, one has
$U^{-} U = \alpha^{-2} U^2 = \alpha^{-2} W W^* W W^* = \alpha^{-1} U$
since $W^* W = \alpha I$.
\\

\noindent
{\bf Proof of Corollary  \ref{cor:lin_unbiased_risk}. }
Note that
$$
\Tr [\Ga U \Ga U^{-}] = \Tr [U U^{-} + (I_N-\Ga) U (I_N-\Ga)U^{-} - 2 (I_N-\Ga) U U^{-}]
$$
Now, to  prove \fr{lin_shrink}, replace $\theta \theta^*$ by $y y^* - \sigma^2 U$ in \fr{lin_shrink_explicit}
and observe that
\be \label{trace_ident}
\Tr [U U^{-}] = \Tr [W W^* (W^{+})^* W^{+}] = \Tr [ (W^{+} W)^*\, W^{+} W] = n
\ee
since $W^{+} W = I_n$. In order to obtain \fr{lin_shrink},
replace $U^{-}$ with $ \alpha^{-2} U$.
\\

\noindent
{\bf Proof of Theorem  \ref{th:risk_genframes}.  }
First, let us show that under conditions of Theorem \ref{th:risk_genframes},
one has
\be \label{main_formula}
\EE [(\hat{\theta} - \theta) (\hat{\theta} - \theta)^* ] = \sig^2 U + \EE [g(y) g^*(y)]
+ 2 \sig^2 U \EE [Z].
\ee
To this end, note that
$$
\EE[\hat{\theta} - \theta)(\hat{\theta} - \theta)^*] =
\EE[(y - \theta)(y-\theta)^* + g(y) g^*(y)] + 2 \EE[(y-\theta) g^*(y)] \equiv
\Omega_1 + 2 \Omega_2.
$$
Here $\Omega_1 = \sig^2 U + \EE[g(y) g^*(y)]$ and, due to representations $y = Wx$ and $\theta = W f$, the
expression for $\Omega_2$ may be written as
$\Omega_2 = W\ \EE [(x - f) g^*(Wx)].$

Denote $C_\sig = (2 \pi \sig^2)^{-n/2}$ and observe that $Q= \EE [(x - f) g^*(Wx)]$
is the $n \times N$ matrix with components
\beqns
Q_{l j} & = & \EE [(x_l - f_l) g_j (Wx)] = C_\sig \int  \cdots \int (x_l - f_l) g_j (Wx)
\exp \lkr -\|x - f \|^2/2 \sig^2 \rkr  dx\\
& = & - C_\sig \sig^2  \int  \cdots \int  g_j (Wx)\
d_l (\exp \lkr -0.5\,  \sig^{-2}\ \|x - f \|^2  \rkr)  dx_1 \cdots dx_{l-1} dx_{l+1} \cdots dx_n\\
& = &   C_\sig \sig^2  \int  \cdots \int \frac{\partial } {\partial x_l} [g_j (Wx)]\
\exp \lkr -\|x - f \|^2/2 \sig^2 \rkr  dx
= \sig^2\ \EE \lkv \frac{\partial } {\partial x_l} g_j (Wx) \rkv.
\eeqns
In the expression above, we denoted differential with respect to $x_l$ by $d_l$ and used integration by parts.

Applying the chain rule, derive that
$$
\frac{\partial } {\partial x_l} [ g_j (Wx)] = \sum_{i=1}^N \frac{\partial } {\partial y_i} [ g_j (y)] W_{il}
= \sum_{i=1}^N Z_{ij} W_{il} = (W^* Z)_{lj}.
$$
Therefore,
$$
\Omega_2 = \sig^2\ \EE(W W^* Z)  = \sig^2\, U\ \EE(Z),
$$
which yield expression \fr{main_formula}.

Now, to complete the proof of \fr{risk_gen_frames}, observe that
\beqns
\EE \| \hat{f} - f \|^2 & = & \EE\ \Tr \lkv
(\hat{\theta} - \theta)(\hat{\theta} - \theta)^* (W^{+})^* W^{+} \rkv \\
& = & \EE\ \Tr \lkv \sig^2 U U^{-} + g(y) g^*(y) U^{-} + 2  \sig^2 U Z U^{-} \rkv
\eeqns
and recall that, by formula \fr{trace_ident}, one has $\Tr [U U^{-}] = n$.
Validity of formula \fr{risk_tight_frames}
follows from the fact that, in the case of a tight frame, one has
$U^{-} U = \alpha^{-2} U^2$.
\\

\noindent
{\bf Proof of Corollary  \ref{cor:soft_thresh}.  }
Validity follows directly from Theorem \ref{th:risk_genframes}
and relations \fr{soft_thresh} and \fr{matrixZ}.
\\

{\bf Proof of Theorem  \ref{th:const_weights}.  }
Note that
\beqns
\EE \| \hat{f} - f \|^2 & = & \lkr \sum_{i=1}^m \lam_i \rkr \EE \| \hat{f} - f \|^2
= \sum_{i=1}^m \lam_i \EE \| \hat{f} - \hat{f}^{(i)} + \hat{f}^{(i)} - f \|^2 \\
& = & \sum_{i=1}^m \lam_i  \lkv \EE \| \hat{f} - \hat{f}^{(i)} \|^2 +  \EE \| \hat{f}^{(i)} - f \|^2
+ 2 \EE \lkr \hat{f} - \hat{f}^{(i)} \rkr^* \lkr \hat{f}^{(i)} - f \rkr \rkv.
\eeqns
By direct calculations, it is easy to check that
$$
 \sum_{i=1}^m \lam_i  \lkr \hat{f} - \hat{f}^{(i)} \rkr^* \lkr \hat{f}^{(i)} - f \rkr
= \| \hat{f} \|^2 - \sum_{i=1}^m \lam_i \| \hat{f}^{(i)} \|^2
 = - \sum_{i=1}^m \lam_i \| \hat{f}^{(i)} - \hat{f} \|^2 .
$$
Therefore, changing the order of expectation and summation (due to the fact that the weights
are data-independent) and using identity above, we derive
$$
\EE \| \hat{f} - f \|^2   =
\EE  \sum_{i=1}^m \lam_i \lkv \| \hat{f}^{(i)} - \hat{f} \|^2  + \| \hat{f}^{(i)} - f \|^2
- 2 \| \hat{f}^{(i)} - \hat{f} \|^2 \rkv,
$$
which completes the proof.
\\

\noindent
{\bf Proof of Theorem  \ref{th:risk_dd_weights}.  }
Applying Theorem \ref{th:risk_genframes}
with $\alpha =m$ and  a block vector $\tilde{g}$  with blocks given by formula \fr{tildeg},
one can write $\Delta$ as
\be \label{del1_del2}
\Delta = m^{-2} \tilde{g}^* (y)\, U\, \tilde{g}(y) + 2 \sig^2 m^{-1}\, \Tr [ U \tilde{Z}] \equiv
\Delta_1 + \Delta_2.
\ee
Here, re-arranging $\tilde{g}^{(i)} (y)$, we obtain
\beqns
\Delta_1 & = & \sum_{i,j=1}^m \lam_i(y) \lam_j(y)  (g^{(i)})^*   U^{(i,j)} g^{(j)}
+ m^{-2} \sum_{i,j=1}^m [1 - m \lam_i(y)][1 - m \lam_j(y)]  (y^{(i)})^*  U^{(i,j)} y^{(j)}\\
& -  & 2 m^{-1}  \sum_{i,j=1}^m \lam_i(y) [1 - m \lam_j(y)] (g^{(i)})^*   U^{(i,j)}  y^{(j)}.
\eeqns
Since $U^{(i,j)} y^{(j)} = W^{(i)} (W^{(j)})^* y^{(j)} = W^{(i)} x = y^{(i)}$ and
$\sum_{j=1}^m (1 - m \lam_j)=0$, the second and the third terms in the last expression are equal to zero and
\beqn \label{delta1}
\Delta_1 & = & \sum_{i,j=1}^m \lam_i(y) \lam_j(y)  (g^{(i)})^*   U^{(i,j)} g^{(j)}.
\eeqn

Now, consider $\Delta_2$. Note that $\tilde{Z}$ is a block matrix with blocks $\tilde{Z}^{(i,j)}$
which, with the help of the product rule, can be presented in the  form
\beqns
\tilde{Z}^{(i,j)} & = & \nabla_{y^{(i)}} (\tilde{g}^{(j)})^*  (y)
= m [ \nabla_{y^{(i)}} \lam_j (y)] (\hat{\theta}^{(j)})^*
+ m \lam_i (y) [ I_n + Z^{(i,i)} ] \II(i=j) - I_n \II(i=j).
\eeqns
Substituting the last expression into $\Delta_2$ in \fr{del1_del2} and recalling that
$U^{(i,i)} = I_n$, we arrive at
\beqn
\Delta_2 & = & 2 \sig^2 m^{-1} \sum_{i,j=1}^m  \Tr[U^{(j,i)} \tilde{Z}^{(i,j)}]
=  2 \sig^2  \sum_{i,j=1}^m  \Tr \lkv U^{(j,i)} [\nabla_{y^{(i)}} \lam_j(y)] \  (\hat{\theta}^{(j)})^*  \rkv
\label{delta2}\\
& + & 2 \sig^2  \sum_{i =1}^m \lam_i (y)\ \Tr[I_n + Z^{(i,i)}] -
2 \sig^2 n. \nonumber
\eeqn
Now, interchange $i$ and $j$ in the first term of $\Delta_2$, and also
note that
$$
 \sum_{i =1}^m \lam_i (y)\ \Tr [I_n + Z^{(i,i)}] = n +  \sum_{i =1}^m \lam_i (y)\ \Tr [Z^{(i,i)}].
$$
To complete the proof, combine \fr{del1_del2}, \fr{delta1} and \fr{delta2}.
\\

\noindent
{\bf Proof of Lemma \ref{lem:SURE_deriv}.  }
Observe that
$\nabla_y [g^*(y) g(y)]$ is a column vector with components
$$
\frac{\partial}{\partial y_l} [g^*(y) g(y)] = 2\, \sum_{k=1}^n
\frac{\partial  g_k  (y)}{\partial y_l} \ g_k(y).
$$
Hence,  $\nabla_y [g^*(y) g(y)] = 2 [ \nabla_y g^* (y) ] g(y).$
Similarly,  $\nabla_y \Tr \lkv \nabla_y g^* (y) \rkv$ is a column vector with components
$$
d_k(y) = \frac{\partial}{\partial y_k}\, \lkv \sum_{l=1}^n \frac{\partial  g_l  (y)}{\partial y_l} \rkv,
$$
which coincides with \fr{d_components}.
\\

\noindent
{\bf Proof of Corollary \ref{cor:Delta0_expo_weights}.  }
Denote
$$
\Psi (y) = \sum_{l=1}^m  \pi_l \exp \lkr - \beta \eta_l \rkr,
$$
so that, $\log(\lam_i (y)) = \log(\pi_i) - \beta  \eta_i (y^{(i)})  - \log (\Psi (y))$.
Then,
$\nabla_{y^{(j)}} \lam_i (y) = \lam_i(y)\  \nabla_{y^{(j)}} [\log(\lam_i (y))]$
where
$$
\nabla_{y^{(j)}} [\log(\lam_i (y))] = - \beta\ \nabla_{y^{(j)}}  [\eta_i (y^{(i)})]
+ \beta\ \sum_{l=1}^m \lam_l (y) \nabla_{y^{(j)}} [\eta_l (y^{(l)})].
$$
Taking into account that $ \nabla_{y^{(j)}} [\eta_i (y^{(i)})] =0$ if $i \neq j$,
we derive
$$
\nabla_{y^{(j)}} \lam_i (y) =   \beta \lam_i (y) \lkv \lam_j(y)  \nabla_{y^{(j)}} [\eta_j (y^{(j)})]
-  \nabla_{y^{(j)}} [\eta_i (y^{(i)})] \II(i=j) \rkv.
$$
Now, to complete the proof of \fr{DD1_form2}, recall from Theorem \ref{th:risk_dd_weights} that
$$
\Delta_0 = 2 \sig^2 \sum_{i,j=1}^m   [ \nabla_{y^{(j)}} \lam_i (y)]^*\  U^{(j,i)}\ \hat{\theta}^{(i)}
$$
and insert the expression for $\nabla_{y^{(j)}} [\lam_i (y)]$ into $\Delta_0$.
To show validity of \fr{DD1_form1}, note that
\begin{align*}
& \sum_{i,j=1}^m \lam_i(y) \lam_j(y) \lkv \nabla_{y^{(j)}} \eta_j (y^{(j)}) \rkv^*\ U^{(j,i)}\ \hat{\theta}^{(i)}   = \\
& \sum_{j=1}^m \lam_j(y) \lkv \nabla_{y^{(j)}} \eta_j (y^{(j)}) \rkv^* W^{(j)}
\sum_{i=1}^m \lam_i(y) (W^{(i)})^*\   \hat{\theta}^{(i)}      =\\
& \sum_{j=1}^m \lam_j(y) \lkv \nabla_{y^{(j)}} \eta_j (y^{(j)}) \rkv^* W^{(j)} \hat{f}. &
\end{align*}

\section*{Acknowledgements}

Marianna Pensky was supported in part by National Science Foundation
(NSF), grant  DMS-1106564.

\end{document}